\documentclass[aps,twocolumn,10pt,showpacs,amsmath,amssymb,superscriptaddress]{revtex4-1}

\usepackage{graphicx}
\usepackage{color}
\usepackage{subfigure}
\usepackage{dcolumn}
\usepackage{bm}
\usepackage{epsf}
\usepackage{amsmath,amstext,amssymb}
\usepackage{enumitem}
\usepackage{hyperref}
\usepackage[utf8]{inputenc}
\usepackage{epstopdf}
\usepackage{appendix}
\usepackage{upgreek}

\PassOptionsToPackage{numbers,sort&compress}{natbib}
\usepackage{hyperref} 

\newcommand{\be}{\begin{equation}}
\newcommand{\ee}{\end{equation}}
\newcommand{\bea}{\begin{eqnarray}}
\newcommand{\eea}{\end{eqnarray}}
\newcommand{\bra}[1]{\left\langle#1\right|}
\newcommand{\ket}[1]{\left|#1\right\rangle}

\begin{document}

\title{Spin and motion dynamics with zigzag ion crystals in transverse magnetic field gradients}

\author{J.~Welzel}
\author{F.~Stopp}
\author{F.~Schmidt-Kaler}
\affiliation{Institut f\"ur Physik, Universit\"at Mainz, Staudingerweg 7, 55128 Mainz, Germany}

\begin{abstract}
We investigate the dynamics of ion crystals in zigzag configuration in transverse magnetic field gradients. A surface-electrode Paul trap is employed to trap $^{40}$Ca$^+$ ions and features submerged wires to generate magnetic field gradients of up to 16.3(9)~T/m at the ions position. With the gradient aligned in the direction perpendicular to the axis of weakest confinement, along which linear ion crystals are formed, we demonstrate magnetic field gradient induced coupling between the spin and ion motion. For crystals of three ions on their linear-to-zigzag structural transition we perform sideband spectroscopy upon directly driving the spins with a radiofrequency field. Furthermore, we observe the rich excitation spectrum of vibrational modes in a planar crystal comprised of four ions. 
\end{abstract}

\maketitle
Quantum systems with tunable interactions and individual manipulation and measurement provide a suitable platform for quantum simulations. Internal quantum states of trapped ions allow for encoding of two-level systems, i.e. spin-$\tfrac{1}{2}$ systems, which can be manipulated and read out at high fidelities~\cite{HAEFFNER2008}. In a Paul trap, $n$ ions can be stored in the form of linear or two-dimensional Coulomb crystals, which exhibit $3n$ collective modes of vibration. To couple internal spin state and external motion, eventually tailoring a spin-spin interaction, electromagnetic radiation with high field gradients is employed. Forces from focused laser beams have been used in such way, for the entanglement of several particles~\cite{MONZ2011,GAEBLER2016}, or alternatively microwave fields. The latter promises a reduction of technical complexity~\cite{MINTERT2001,chiaverini2008laserless,weidt16} to mediate spin interactions. Experiments are either using oscillating~\cite{ospelkaus11,HARTY2016} or  static~\cite{weidt16} magnetic field gradients to induce coupled spin-motion dynamics. To achieve the required static field gradients, approaches with permanent magnets~\cite{khromova12,lake15,kawai17} or electro-magnets~\cite{kunert2014} have been implemented. The resulting long-range spin interactions allows for designing spin-boson~\cite{Porras2008,Wall2017} or spin-spin~\cite{Kim2010,Lanyon57} models, with relevance to our understanding of quantum magnetism~\cite{Jurcevic2014,Islam583}, for studies of non-linear dynamics~\cite{GESSNER2014,LEMMER2015} and quantum phase transitions~\cite{JOHANNING2009}. Experimental highlights are simulating phase transitions in interacting linear spin systems~\cite{ZHANG2017,JURE2017}. Eventually extending from a linear to two-dimensional self-assembled spin arrays~\cite{BRI12,BER11} or even fully controlled geometries of interacting spins~\cite{MIE16,KUM16} opens up investigations of interesting quantum phases in systems with frustrated interactions. 

In this work, we store small planar ion crystals and employ a static magnetic field gradient pointing along the weaker one of the two radial axes of the trap, i.e. the direction perpendicular to the trap axis. This allows for a simultaneous manipulation of the spins of all ions within a chain, and therefore enables the investigation of magnetic field  gradient-induced excitations of vibrational modes in crystals at the crossover from a linear to a zigzag structure. After briefly recapitulating the interaction between an ion and an electromagnetic field in the presence of a strong magnetic field gradient, we present our experimental apparatus. First we characterize it with measurements on a single ion, and then present results on the excitation of three and four ion zigzag crystals. With the aim of performing quantum simulations in spin-boson systems, realized in larger planar ion Coulomb crystals, we discuss current technological limitations of the approach and propose how these could be overcome. 

\begin{figure}[!tb]
\centering
\includegraphics[width=0.99\columnwidth]{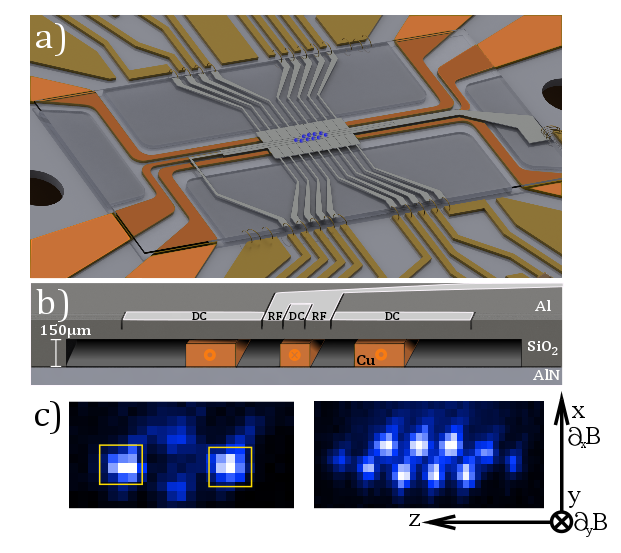}
\caption{a) Sketch of the employed surface electrode ion trap made from fused silica (SiO$_2$, dark grey) and coated with a layer of 0.4~$\upmu$m aluminum, with two rf, one central dc rail and further 2$\times$9 segments for confinement with submerged current wires (orange), which allow for the generation of a transverse magnetic gradient. A surrounding ground plane is not displayed for illustration purpose. b) Cross section of the chip with $50~\upmu$m deep and 10~$\upmu$m wide trenches (Translume Inc.) between the trap electrodes (light grey) and a cut-out at the backside to fit in three current carrying wires (orange) to generate a magnetic quadrupole field at the ion position. c) Fluorescence image of four and 10 ion crystals in planar configuration. Regions of interest (ROI) for counting photons selectively from one, or a selected number of ions.}
\label{figtrap}
\vspace{-10pt}
\end{figure}

The Hamiltonian of an ion of mass $m$ and two internal states $\{\ket{\uparrow}$, $\ket{\downarrow}\}$  in a harmonic potential described by a trap frequency $\omega_\text{t}$ is given by  ${\hbar \omega_\text{t} a^\dagger a+\frac{\hbar}{2}\omega_\text{a}\sigma_z}$, where $\omega_\text{a}$ is the angular frequency of the atomic transition, $a$, $a^\dagger$ the bosonic ladder operators and ${\sigma_z =\ket{\uparrow}\bra{\uparrow}- \ket{\downarrow}\bra{\downarrow}}$. Driving this two-level system rotates the spin $\sigma^{(+)}=\ket{\uparrow}\bra{\downarrow} $ in combination with excitation of motional quanta. The strength of interaction is characterized by the Rabi frequency $\Omega$ and detuning of the driving field from the two-level transition frequency $\delta = \omega_\text{f}-\omega_\text{a}$. Depending on this detuning either carrier transitions ($\delta=0$), the red ($\delta=-\omega_\text{t}$) or the blue  ($\delta=+\omega_\text{t}$) sideband transitions are excited. While in the first case, only the spin state is rotated, in the other cases this goes along with the creation or annihilation of motional quanta. This dynamics is determined by the Lamb-Dicke factor $\eta = \sqrt{\hbar k_{\text{eff}}^2/2 m\omega}$,  that depends on the effective wave vector of the incident driving field $k_{\text{eff}}$, its component in direction of the eigenvector of the ions's oscillation in the trap at frequencies $\omega_{x,y,z}$~along the trap directions~\cite{HAEFFNER2008}. In case of an ion crystal with $n$ ions, there are $3n$ oscillatory modes. Now, $\eta$ becomes specific for every mode and depends on its eigenfrequency $\omega_n$ and the directions of its eigenvectors. Especially in planar ion crystals, the directions of eigenvector and frequencies of modes allow for tailoring complex spin interactions.   

For ions in Paul traps, with mode frequencies $\omega_n$ of a few MHz, and in case of a typical magnetic dipole transition in the radiofrequncy range, resulting in low spin-motion coupling rates and direct magnetic spin-spin interactions in the mHz regime~\cite{KOTLER2014}. Applying a strong magnetic field gradient, however, makes the transition frequency $\omega_\text{a}(x,y,z)$ depend on the ion position in the trap. Thus, a momentum kick is contributed to an ion when a spin flip takes places. A corresponding small displacement from the equilibrium position in a gradient, here directed along the radial x-axis of the trap, results in an additional term in the Hamiltonian $\frac{\hbar}{2}\left.\partial_x \omega_\text{a}(x)\right|_{x_0}\xi\sigma_z$\cite{WUNDERLICH2002}. For a crystal of ions we finally obtain an effective Lamb-Dicke factor of
\begin{equation}
\label{eqEtaeff}
\tilde\eta_{nl}  = S_{nl}\frac{\mathit\Delta m_\text{S} \hspace{1mm} g \hspace{1mm} \mu_\text{B} }{\sqrt{\hbar m\omega_n^3}} \left.\frac{\partial B(x_l)}{\partial x_l}\right|_{x_{0,l}},
\end{equation}
with the difference in the projection of the total angular momentum represented by $\mathit\Delta m_\text{S}$, Land\'e factor $g$, Bohr magneton $\mu_\text{B}$, the absolute value of the magnetic field $B$, and where $S_{nl}$  are normalized displacements for the $n$th collective mode of vibration and the $l$th ion in the crystal. Taking into account typical experimental conditions for ion traps we find that magnetic  field gradients on the order of 10 to 100~T/m are required to drive dynamics on a ms-timescale. 

We use $^{40}$Ca$^+$ confined in a planar Paul trap, which features trapping electrodes consisting of a  $415~$nm thin layer of aluminum evaporated on a fused silica (SiO$_2$) trap chip, see  Fig.~\ref{figtrap}(a). We employ a five-wire design~\cite{Chiaverini05} with asymmetric rf electrodes. The width of the electrodes as seen in FIG.~1 b) is 700,~120,~100,~100,~700~$\upmu$m, respectively. This results in a trapping height of $96.9~\upmu$m above the aluminum surface~\cite{House08}. Segmented outer dc electrodes have a lenght of 430~$\upmu$m along the trap z-axis direction. The ponderomotive potential is generated in the $xy$ plane by a drive frequency of $\Omega_\text{drive}=2\pi\times31.5~$MHz. Crystals with up to 10 ions, see Fig.~\ref{figtrap}(c) have been trapped using an rf peak-to-peak amplitude of $\sim$100~V resulting in a trap depth of about 35~meV. Typical secular trap frequencies are $2\pi\times(2.9,~1.7,~0.89)~$MHz, with the lowest frequency corresponding to the trap z-axis. The trap features a $150~\upmu$m deep cut-out on the backside to allow for the current carrying Cu-wires to be arranged as close to the ions as possible. Wires pointing parallel to the trap z-axis underneath the trap chip surface are etched from a $127~\upmu$m thick copper layer placed on a AlN chip carrier; the wire center to trapped ion distance is close 285$\upmu$m, see Fig.~\ref{figtrap}(b)

Zeeman sublevels of the $^{40}$Ca$^+$ S$_{1/2}$ ground state, $m_\text{S}$=+1/2~$\equiv\ket{\uparrow}$ and $m_\text{S}$=-1/2~$\equiv\ket{\downarrow}$ in a magnetic field of $3.5\times10^{-4}$~T perpendicular to the trap surface generated by external coils, split up in frequency by $2\pi\times10~$MHz. To drive the spin flip transition, we apply an rf current to the central current wire. The magnetic field gradient is generated by all three current wires: a magnetic quadrupole field forms at the position of the ions by applying static currents of +5.8,~-4.8,~+8.3~A, respectively. The asymmetric setting of currents i required for aligning the magnetic quadrupole to the trap potential, and we achieve a gradient of $16.3(9)~$T/m at the ions position. We characterized the gradient by displacing the ion in radial direction from the trap center and measuring its Zeeman splitting~\cite{WARRING2013}.

\begin{figure}[tb]
\centering
\includegraphics[width=1.0\columnwidth]{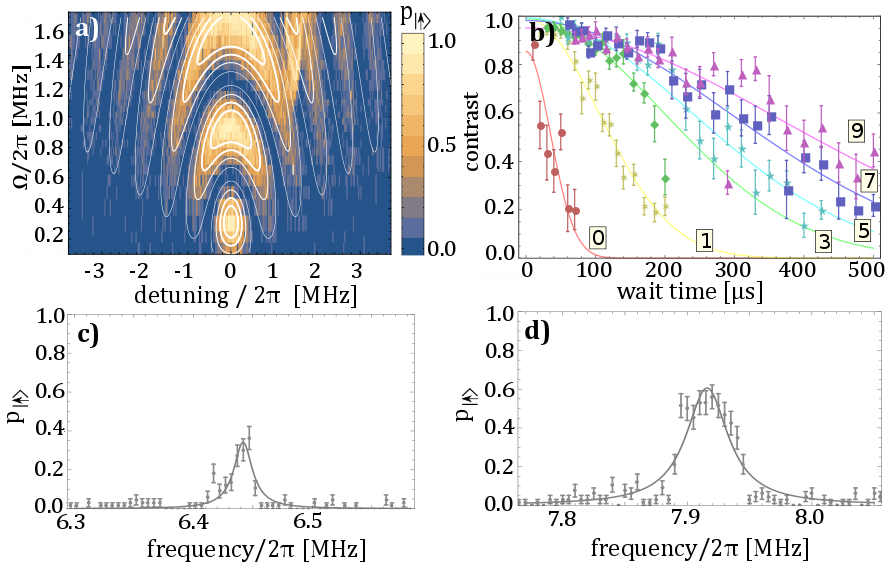}
\caption{Rf spectroscopy with a single ion. a) Excitation probability $\ket{\uparrow}$ on the carrier transion as function of strength and detuning of the rf field, with pulsetime  $\tau=1.7~\upmu$s and {\it r}~$=30$. The experimental data for the Rabi frequency is inferred from rf pulse length scans.  A feature appears for strong drive near a detuning of $+2\pi\times$1.5~MHz, thus at 14.3~MHz absolute frequency, and does not dependent on the external magnetic field strength. We attribute it to a non-linear resonance~\cite{Alheit1996}, a subharmonic of the  frequency differences of the trap drive and the radial sideband, at $\Omega_\text{Paul} - \omega_{\text{rad}}$ = $2\pi\times$14.3~MHz. To avoid spectral overlap with motional sidebands we decreased for all later measurements the offset magnetic field, thus the carrier center frequency from before $2\pi\times$12.7~MHz to then $\leq$10.0~MHz. The theoretical calculation of Rabi flops are indicated by white contour lines for 5\% (thin), 25\%, 50\%, and 75\% (thick) excitation. b) Spin-echo measurements for different numbers of rephasing pulses. c) and d) Rf spectroscopy of radial red sideband transitions. We scan the frequency over the y- and x-mode, respectively and we observe the spin exciation with $\tau =500~\upmu$s, {\it r}~$=66$.}
\label{fig1ion}
\vspace{-15pt}
\end{figure}

An experimental sequence starts with (i) Doppler cooling, (ii) optical pumping, optionally (iii) coherent excitation of the motional state. For this initialization of the ion crystal~\cite{Uli2009}, we use the 4S$_{1/2}$ to 4P$_{1/2}$ transition for Doppler cooling resulting in a mean phonon number of $n_{\text{rad}} \sim$20. Because of the constraint on the magnetic field direction we have chosen a frequency-selective optical pumping scheme yielding 0.98(2) fidelity of $\ket{\downarrow}$. Next, we continue with a (iv) sequence of radiofrequency pulses for spectroscopy or coherent manipulation. Note that the ions in a linear crystal experience the same magnetic field and radial magnetic field gradient. For individually addressing single ion spins in a linear crystal~\cite{PILTZ2014} we may tilt a crystal with respect to the trap axis, following a procedure we developed recently in a multi-layer Paul trap~\cite{KAUFMANN2017} or use a planar ion crystal which extends in radial direction. As a result, ions experience different magnetic field strengths and undergo different excitation. The readout of ions relies on (v) shelving followed by a (vi) fluorescence detection. For this, we transfer the population of the $\ket{\downarrow}$ level to the metastable 3D$_{5/2}$ $m_\text{S}$=-3/2  and $m_\text{S}$=+1/2 using a rapid adiabatic passage pulse~\cite{Wunderlich2006}. The fluorescence emitted upon resonant excitation of the 4S$_{1/2}$ to 4P$_{1/2}$ transition is detected on an EMCCD camera, and recorded in a single  or multiple regions of interest (ROIs) that are set to selectively detect individual ions in the crystal, see inset of Fig.~\ref{figtrap}(a). After a number of repetions {\it r} we determine the probability $P_{\ket{\uparrow}}$.  

\begin{figure}[b]
\centering
\includegraphics[width=1\columnwidth]{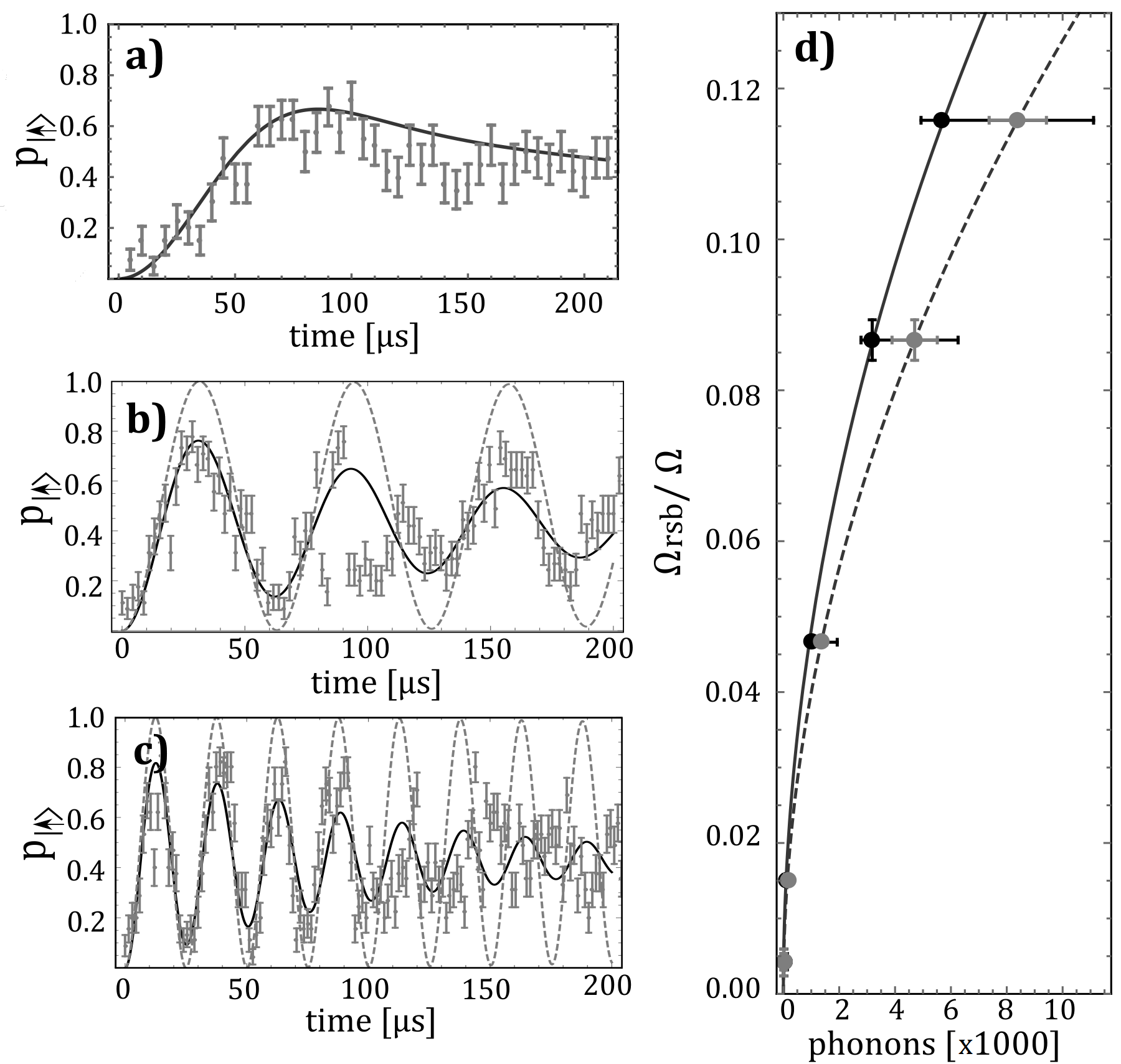}
\caption{Rabi oscillation on the red sideband, at 2$\pi\times$1.05~MHz, for a single ion in a) a thermal state with a fit assuming a mean phonon number of $20.9(2.5)$. b) Rabi oscillation for a coherent state of $\bar n_{\text{coh}}=1360(180)$ phonons and c) 8400(1000). An exponential decay with $\tilde\tau$ = 130(30)~$\upmu$s is taken into account for the fit (solid line). d) Determination of the effective Lamb-Dicke factor on the red sideband enhancement from coherent excitation: Rabi frequency data $\Omega_{\text{rsb}}/ \Omega =\eta \sqrt{n}$  for the the spin transition are fitted with a calculated $\eta$ (grey, dotted) or scaled with mean phonon numbers, extracted from Rabi frequency data of the optical transition (black, solid). The error in the determination of $\bar n_{\text{coh}}$ comes from an uncertainty of the laser direction, and magnetic field gradient direction, with respect to the eigenvector of the radial mode. In (b,c,d) a radial frequency of  $\omega_{\text{rad}}/2\pi = 2.02$~MHz was used.}
\label{figcoherentState}
\end{figure}

We characterize the radiofrequency excitation strength and the coherence with a single ion. Fig.~\ref{fig1ion} shows carrier and the red/blue motional sideband excitation. Under typical experimental conditions we drive the carrier transition with a Rabi frequency of $2\pi\times1\text{MHz}$. To optimize the excitation strenght of the motional sideband, we employ the trap control voltages and align the direction of the lower frequency radial mode (x) along the magnetic field gradient direction. The second radial mode (y), oriented almost perpendicular, can be tuned to show a 20~times weaker coupling. To determine an upper bound for the component of the magnetic field  gradient pointing along the trap axis we employ a two ion crystal and vary the inter-ion distance between 5.6(4) and 12.1(3)~$\upmu$m. We measure the dependence of the qubit frequency on the ion distance and infer a magnetic field gradient less than  0.02~T/m, resulting in a negligible axial sideband excitation with $\eta \leq 10^{-5}$. 

The first red sideband  Rabi frequency depends on the phonon number $n$ in the mode as $\Omega_{\text{sb}}(n)=\Omega e^{-\eta^2/2}\eta L_n^1(\eta^2)n^{-1/2}$~\cite{wineland79} where $L_n^m(x)$ denote associated Laguerre polynomials. For a magnetic field gradient of 16.3(9)~T/m and  a mode frequency of $2\pi\times2.02(1)~$MHz we calculate an effective Lamb-Dicke factor of $\eta=0.00126(7)$. Rabi oscillations result from the summation with the phonon number distribution $p_n$ as $p_{\ket{\uparrow}} =\sum{p_n \sin^2\left[(1/2) \Omega_{\text{sb}}(n)t\right]}\label{eqp}$. For an ion in a thermal state after Doppler cooling the dynamics of the red sideband is shown in Fig.~\ref{figcoherentState}(d). To increase the effective Rabi frequency $\Omega_{\text{sb}}$ we inject an alternating resonant electric field to excite the radial mode coherently. Then we drive Rabi oscillations for different power levels of electric excitation, see Fig.~\ref{figcoherentState}(b, c), and obtain the mean phonon number of the coherent distribution, with $\bar n_{\text{coh}}$= 8400(1000) and 1360(180), respectively. A fast red sideband transition was achieved in $12~\upmu$s. To cross-check the results of $\bar n_{\text{coh}}$, we drive Rabi oscillations on the optical S$_{1/2}\leftrightarrow$D$_{5/2}$ transition, and its variation with the same injected electric fields and obtain $\bar n_{\text{coh}}=5600(^{+5500}_{-700})$ and $990(^{+940}_{-120})$, respectively, see Fig.~\ref{figcoherentState}(d)~\cite{home16}. This way, we extract $\eta = 0.0015(^{+1}_{-7})$ from the fit, which agrees within errors with the calculated value. 

To measure the coherence time we use a single ion, apply a $\pi$/2 carrier pulse, followed by a wait time, and a second $\pi$/2 pulse. Ramsey fringes are recorded, when scanning the phase of the second pulse. The fringe contrast is fitted by a Gaussian decay with a $1$/e-time of $\sim50~\upmu$s, see Fig.~\ref{fig1ion}(b). For an odd number {\em l} of $\pi$ spin-echo refocussing pulses, we find that the measured coherence time scales with {\em l}$^{ 0.64(13)}$  and for {\em l} $=9$ we reach $\sim0.5$ ms.  Such power law indicates a Lorentzian noise spectrum~\cite{BARGILL2013} which in our case is fully dominated by fluctuations of currents $\leq 2$~mA$_\text{rms}$ to generate the magnetic field gradient. Trap control voltages are sufficiently stable such that the ion crystal is kept aligned within the gradient field within a small fraction of the radial wave packet size of 39(3)~nm.

\begin{figure}[bt]
\centering
\includegraphics[width=0.99\columnwidth]{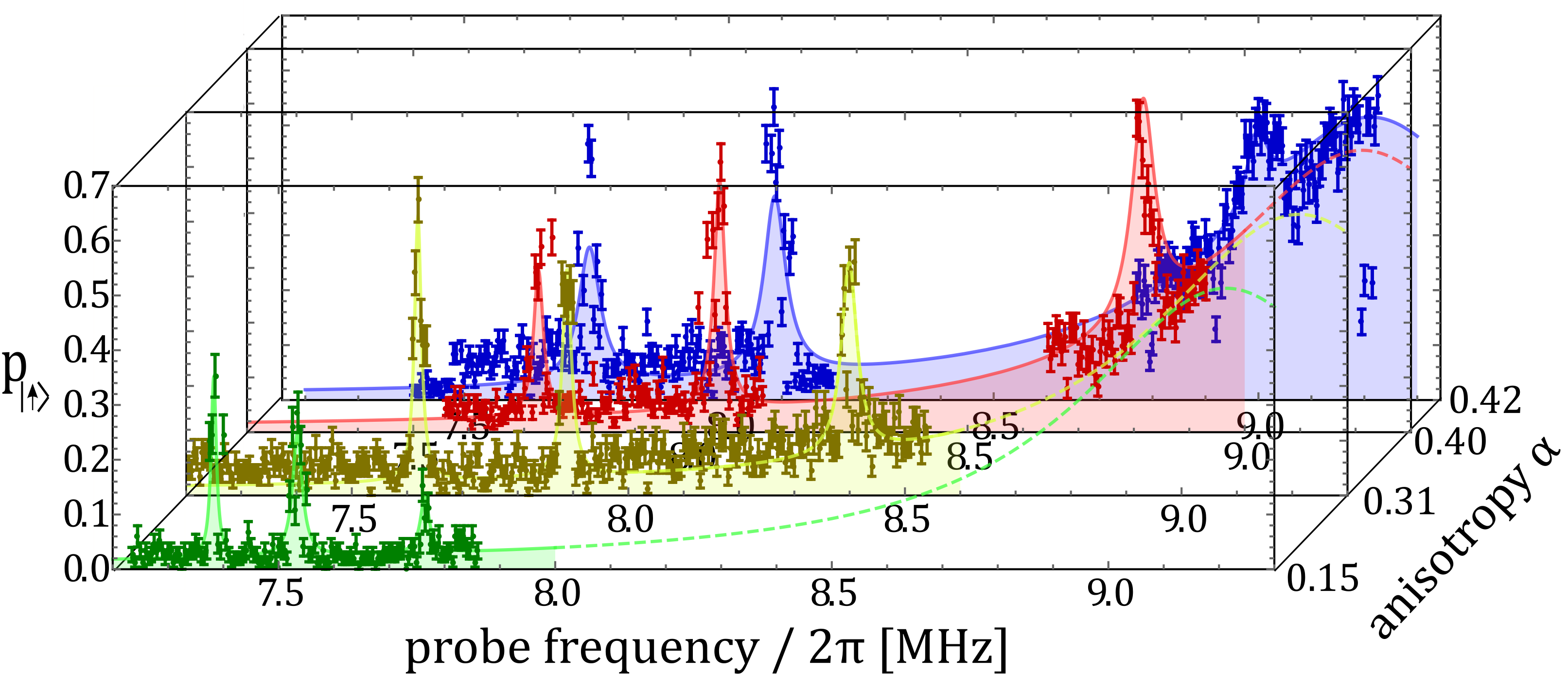}
\caption{Spectrum of three ion crystal taken with ROI on the outer ion, with four different sets of trap frequencies with $\alpha=\{$0.15, 0.31, 0.40, 0.42$\}$. For~$\alpha=0.4$ the zigzag mode shifts to $2\pi\times350~$kHz (red points). At~$\alpha=0.42$ the critical point is already passed  (blue points). The number of repetitions was ${\it r}=150$, $\tau=1.75~$ms, and carrier transition was at $2\pi\times$9.2~MHz.}
\label{fig3ionen}
\vspace{-15pt}
\end{figure}

We investigate spin transitions in the vicinity of a structural transition of a small ion crystal by adjusting the anisotropy parameter $\alpha=(\omega_{\text{ax}}/\omega_{\text{rad}})^2$ to generate  the desired configurations. For small anisotropy ions arrange in linear configuration along the trap axis. For $\alpha$ above the critical value $\alpha_ {\text{crit}}$ ions are forming a zigzag or planar crystal~\cite{birkl92,fishman08,KAUFMANN2012}. In case of a  three-ion crystal we drive the spectroscopy sequence (i) to (v) recording fluorescence only on one of the outer ions, see Fig.~\ref{fig3ionen}. All three modes with the eigenvectors (1,1,1)/$\sqrt3$, (1,0,-1)/$\sqrt2$ and (1,-2,1)/$\sqrt6$ containing radial x-direction appear. Increasing the value of $\alpha$ towards $\alpha_ {\text{crit}}$, here 0.416(1), eigenfrequencies become smaller such that the observed resonances approach the carrier and getting more pronounced as their effective Lamb-Dicke factors $\eta$ are increasing. The spectrum for $\alpha$ = 0.42 is taken already beyond the critical point with a zigzag crystal extending in the radial direction by about 1.2~$\upmu$m. The lowest sideband mode near a drive frequency of $2\pi\times9$~MHz corresponds to a zigzag vibration of the crystal, with an eigenfrequency of $2\pi\times215(14)$~kHz, consistent with the calculated value of $2\pi\times225$~kHz.      

\begin{figure}[tb]
\centering
\includegraphics[width=0.99\columnwidth]{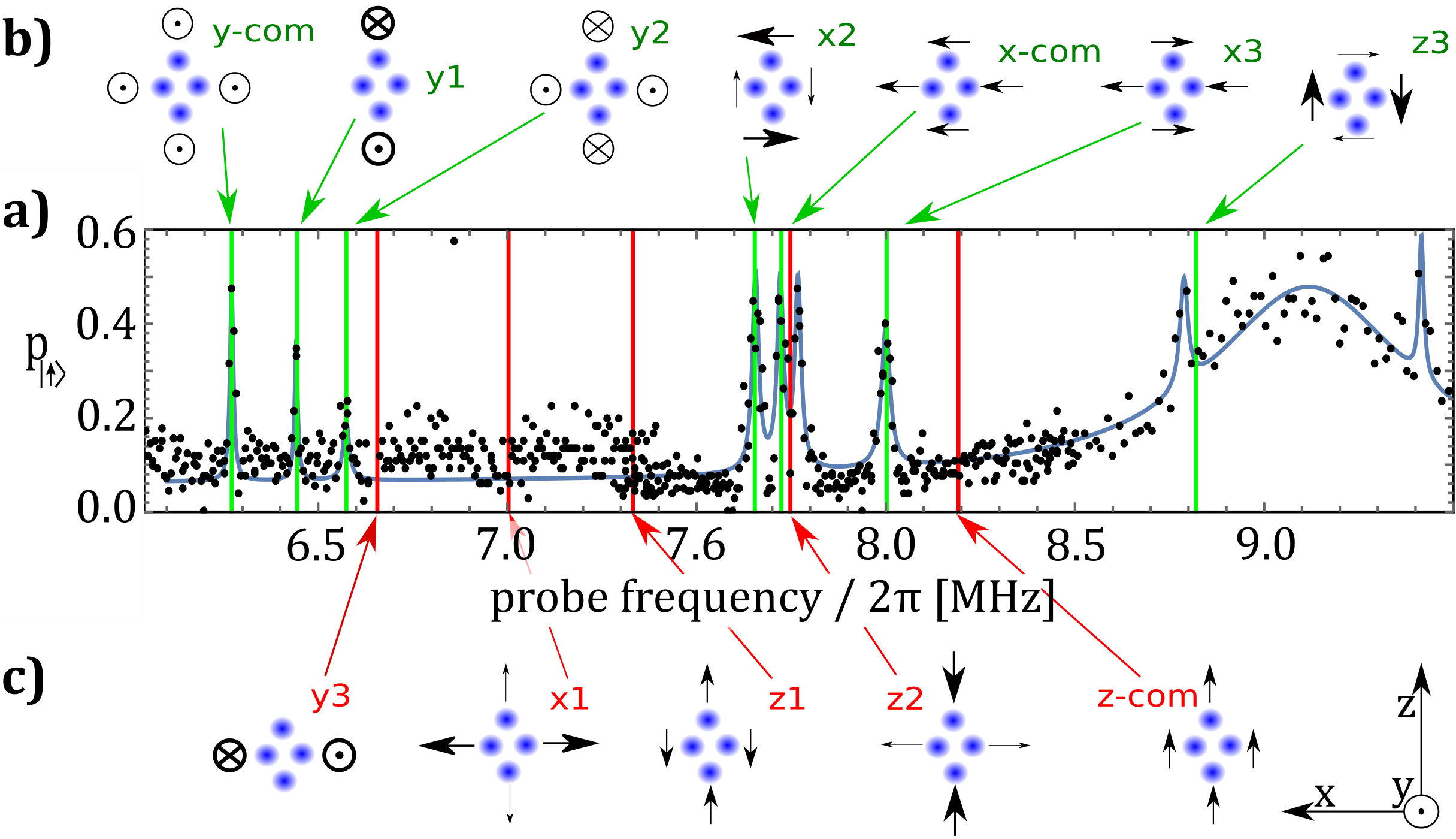}
\caption{a) Spectrum of four ion planar crystal with {\it r}~$=66$, $\tau=5$~ms, and a Rabi frequency of 2$\pi\times 450~$kHz. Error bars are omitted for clarity. Vertical lines indicate the calculated position of the eigenmodes from a pseudo potential approximation with an ac-Zeeman correction. Visualization of eigenmodes b), c): The relative strength of the motion is represented by the arrow size, (x, y, z) denote the predominant direction of motion, (1,2,3) indicate the highest to lowest mode frequency. b) Eigen modes which are observed in the spectrum, since their eigenvectors of ROI ions have non-vanishing projection along the magnetic field gradient direction. c) Modes, which are not excited, as the oscillation direction is perpendicular to the direction of the magnetic field gradient. The z${2}$ mode excitation is an exception, we assume mixing with the strong nearby x-com mode.}
\label{fig4ionen}
\vspace{-20pt}
\end{figure}

We turn now to the basic case of a planar crystal with four ions, see Fig.~\ref{figtrap}(c). We initialize the two ions which are placed on the rf node of the trap. After an rf pulse of 5~ms their spin state is measured using fluorescence detection (within ROIs). Applying the spectroscopic sequence we investigate the 4$\times$3 eigenmodes of common vibration. The magnetic field gradient couples only vibrational modes to the rf transition, which feature ion oscillations in the radial direction, i.e. parallel to the gradient, cf. Fig.~\ref{fig4ionen}.  The radial x-modes, which are aligned with the gradient, show up as resonances x-${\text{com}}$, x${2}$ and x${3}$ (green), while the eigenvectors of x${1}$ do not point along the magnetic field  gradient and are therefore not excited, resonance frequencies indicated in red. As the rf excitation was at high power, saturating the transition with $\Omega_{\text{Rabi}}\sim 2\pi\times450~$kHz we also observe the radial modes with weaker projection y-${\text{com}}$, y${1}$, and y${2}$ (green), while for y${3}$ only ions move which are not in ROIs (red). The axial z-modes do not appear, except z${2}$, which is in close spectral vicinity to the x-${\text{com}}$ mode. Here, we assume that the strong rf drive field induces a mixing process between x-${\text{com}}$ and z${2}$, resulting in dressed modes which show up in the spectrum. As ions are not all located on the radiofrequency node of the trap they undergo micromotion, a precise calculation of eigenfrequencies would require a Floquet-Lyapunov~\cite{KAUFMANN2012}. Because of a sufficiently small Matthieu parameter $q=0.22$ even a pseudo-potential calculation fits the observed resonances well. 

We have shown coupled dynamics of motional and internal spin states with a transition frequency as low as $2\pi\times10$~MHz by employing static magnetic field  gradients created with current-carrying wires placed underneath an ion trap chip. We presented detailed resolved-sideband spectoscopy in linear and planar multi-ion crystals. Knowledge of the eigenmodes and frequencies are paramount for the implementation of quantum simulation and tailoring of interaction strengths between spins. 

In future experiments we plan to significantly improve coherence time and magnetic field gradient (up to 150~T/m) which can be achieved by using a much more advanced current suppy scheme or by permanet magnets placed underneath the trap chip. Slow decoherence may be corrected for by advanced refocusing schemes~\cite{PILTZ13}. Recently fast and robust gate operations have been proposed~\cite{ARRAZOLA2017} and would be in experimental reach. Another interesting direction are studies of phase transitions with competing spin-spin interactions~\cite{Kim2010} implemented on larger planar ion crystals or studies of revealing nonlinear mode couplings~\cite{GESSNER2014,LEMMER2015}

We acknowledge A. Bautista-Salvador and N. Daniilidis for help with the trap fabrication. We thank U. Poschinger and B. Lekitsch for helpful comments and careful reading. We  acknowledge financial support by the DFG through the DIP program Grant No. SCHM 1049/7-1,  the SFB/TR-49 and within the EU-STREP EQuaM. 

\bibliographystyle{apsrev4-1}
\bibliography{ref}

\end{document}